\documentstyle[preprint,eqsecnum,aps,epsf,floats,tighten]{revtex} 

\begin{document}

\preprint{\tighten \vbox{\hbox{hep-ph/9808313} 
		\hbox{} \hbox{DO-TH 98/13} }}

\title{Probing hadron structure and strong interactions with inclusive 
semileptonic decays of $B$ mesons}

\author{Changhao Jin}

\address{School of Physics, University of Melbourne\\
Parkville, Victoria 3052, Australia}

\maketitle

{\tighten
\begin{abstract}%
The study of inclusive semileptonic decays of $B$ mesons is analyzed
from the viewpoint of probing hadron structure and strong interactions.
General formulas for the differential decay rates are given in
terms of the structure functions in arbitrary frame of reference, taking
into account the finite charged lepton mass. These formulas can be used
for structure function measurements.  The behavior of the structure
functions is shown to contain information on the constitution of the $B$
meson and the dynamics of strong interactions. 
Measurements of the structure functions would reveal the existence of a
point-like $b$ quark in the $B$ meson, establish the spin $1/2$ nature of 
the $b$ quark, and identify the $b$ quantum number of the $B$ meson.  These
measurements would determine the $b$ quark distribution function, and allow 
new insight into the nature of confinement. The $b$ quark distribution
function can also be extracted directly through the measurements of the decay 
distributions with respect to the scaling variable $\xi_+$.  
\end{abstract}
}

\newpage

\section{Introduction}
Coloured quarks and gluons as physical constituents of hadrons and quantum 
chromodynamics (QCD) 
as the theory of strong interactions have been generally accepted as
the modern point of view with regard to the structure of hadrons and 
strong interactions.  
Asymptotic freedom of QCD explains Bjorken scaling \cite{bj} observed in deep 
inelastic electron-proton scattering \cite{dis}
before the advent of QCD and allows perturbative 
calculations of strong interactions at high energy or short distance.  
On the other hand, the increase of the strong 
coupling constant of QCD at low 
energy or long distance provides a reason to accept quarks and gluons as
physical constituents of hadrons without directly detecting free quarks and
gluons.  There are a variety of experimental results, which
not only provide confirmations of the quark and gluon structure of hadrons, 
but also lead to incisive precision tests of QCD.  
While the evidence is compelling that QCD is a realistic theory of strong 
interactions, there is still a very incomplete understanding of
the nonperturbative, confining, long-distance sector of QCD,  
which is a basic challenge in describing strong 
interactions, in particular how quarks and gluons bind together to form 
hadrons. 

The $B$ meson is heavy and relatively long-lived \cite{PDG}. This makes it
advantageous to probe hadronic physics.
One powerful way of experimentally investigating the structure of the $B$ meson
and strong interactions is to study its inclusive semileptonic decay.
The structure of the $B$ meson and the strong interaction can be studied with
this underlying weak decay process because the decay rate is modified by the 
strong interaction, which is responsible for quark confinement and gluon 
radiation. 
The power and beauty of inclusive semileptonic decay is that the electroweak
field generated during the semileptonic decay is well understood and leptons
do not interact strongly.  Furthermore, because of
inclusive characteristics of the decay we would expect that the decay rate is 
mainly affected by the initial bound state effect and is insensitive to 
hadronization in the final states, which are summed
over and possess the completeness.
These permit us to probe the structure of the $B$ meson and
strong interactions by means of a known charged weak current. 

While much attention has been focussed on the determination of $|V_{ub}|$
and $|V_{cb}|$ by inclusive semileptonic decays of $B$ mesons,
a wealth of information on the hadron structure and strong interactions
available from these processes deserves intense scrutiny in experiment and 
theory as well.  The purpose of this paper is to discuss
the study of hadron structure and 
strong interactions by means of inclusive semileptonic $B$ decays.  
We would like to show that this study will provide evidence for the quark 
structure of the $B$ meson and 
allow new insight into the nature of confinement.  Of course,   
this study will also be beneficial to
precise determination of the fundamental standard model parameters 
$|V_{ub}|$ and $|V_{cb}|$. 

The outline of this paper is as follows.  We derive in Sec.~II
within the framework of the standard electroweak model
general formulas for the differential decay rates for inclusive 
semileptonic $B$ meson decays in terms of the structure functions.
The derivation is made in arbitrary frame of reference and the finite charged 
lepton mass is taken into account. These formulas are useful
for measurements of the structure functions. 
The behavior of the structure functions is explored in Sec.~III, 
where we also elucidate by measurements of the structure functions what can 
be learned on the constitution of the $B$ meson.
The extraction of the $b$ quark distribution function, to which the structure
functions are related due to the light-cone dominance, is discussed in 
Sec.~IV.  In Sec.~V we conclude.

\section{Structure Functions}
We want to compute the decay rate for 
the inclusive semileptonic decay of the $B$ meson to lowest order in the
weak interaction in arbitrary frame of reference.  
In this inclusive process $\bar B\to X\ell\bar{\nu_\ell}$,
the $B$ meson of mass $M$ with four-momentum $P$ decays
into any possible hadronic 
final state $X$, over which we sum,  via the emission of 
a virtual $W$ boson which materializes as a charged lepton $\ell$ of mass 
$M_\ell$ with four-momentum $k_\ell$ and a massless antineutrino 
$\bar{\nu_\ell}$ with four-momentum $k_\nu$. 
By standard steps we find
the unpolarized decay rate to be proportional to the leptonic tensor 
$L^{\mu\nu}$ times the hadronic tensor $W_{\mu\nu}$:
\begin{equation}
d\Gamma= \frac{G_{F}^2\left|V_{qb}\right|^2}{(2\pi)^5E}L^{\mu\nu}
W_{\mu\nu}\frac{d^3k_\ell}{2E_{\ell}}\frac{d^3k_{\nu}}{2E_{\nu}} \, ,
\label{eq:kga}
\end{equation}
where $E$, $E_\ell$, and $E_\nu$ denote the energies of the $B$ meson, the 
charged lepton, and the antineutrino, respectively. 
$V_{qb}$ is the Cabibbo-Kobayashi-Maskawa (CKM) matrix element,
which is $V_{cb}$ for the $b\to c$ transition induced decay and $V_{ub}$ for
the $b\to u$ transition induced decay.
The leptonic tensor for the lepton pair is completely determined by the
standard electroweak theory since leptons do not have strong interactions: 
\begin{equation}
L^{\mu\nu}= 2(k^{\mu}_{\ell}k^{\nu}_{\nu}+k^{\mu}_{\nu}k^{\nu}_{\ell}-
g^{\mu\nu}k_{\ell}\cdot k_{\nu}+i\varepsilon^{\mu\nu}\hspace{0.06cm}_{\alpha\beta}
k^{\alpha}_{\ell}k^{\beta}_{\nu}).
\label{eq:klepton}
\end{equation}
The hadronic tensor incorporates all nonperturbative QCD physics for the 
inclusive semileptonic $B$ decay. It is 
summed over all hadronic final states and can be expressed in terms of a 
current commutator taken between the $B$ meson states:
\begin{equation}
W_{\mu\nu}= -\frac{1}{2\pi}\int d^4y e^{iq\cdot y}
\langle B\left|[j_{\mu}(y),j^{\dagger}_{\nu}(0)]\right|B\rangle ,
\label{eq:comm2}
\end{equation}
where $q=k_\ell+k_\nu$ stands for the momentum transfer from the $B$ meson to 
the lepton pair and $j_{\mu}(y) = \bar{q}(y)\gamma_{\mu}(1-\gamma_5)b(y)$ is 
the charged weak current.
The $B$ meson state $|B\rangle$ satisfies the standard covariant
normalization $\langle B|B\rangle=2E(2\pi)^3\delta^3({\bf 0})$. 
 
The most general hadronic tensor form that can be contructed is a 
linear combination of $P_\mu P_\nu,\,
P_\mu q_\nu,\, q_\mu P_\nu,\, q_\mu q_\nu,\, 
\varepsilon_{\mu\nu\alpha\beta}P^\alpha q^\beta$, and $g_{\mu\nu}$, 
with coefficients being scalar functions $W_a(\nu,\, q^2)$ of the two 
independent Lorentz invariants,
$\nu\equiv q\cdot P/M$ and $q^2$.  However,   
the combination $P_\mu q_\nu-q_\mu P_\nu$ does not contribute
since $L^{\mu\nu}(P_\mu q_\nu-q_\mu P_\nu)=0$.  Thus the hadronic tensor 
must take the form
\begin{equation}
W_{\mu\nu} = -g_{\mu\nu}W_1 + \frac{P_{\mu}P_{\nu}}{M^2} W_2 
 -i\varepsilon_{\mu\nu\alpha\beta} \frac{P^{\alpha}q^{\beta}}{M^2}W_3
         + \frac{q_{\mu}q_{\nu}}{M^2} W_4 
 + \frac{P_{\mu}q_{\nu}+q_{\mu}P_{\nu}}{M^2} W_5 \, .
\label{eq:exp2}
\end{equation}
Equation (\ref{eq:comm2}) shows that $W_{\mu\nu}^{\ast}=W_{\nu\mu}$,
so $W_a, a=1,\ldots, 5$, are real.
The interesting physics describing the hadron structure and
strong interactions is wrapped up in the five dimensionless real structure 
functions $W_a(\nu,\, q^2), a=1,\ldots, 5$, for the unpolarized processes.

The unpolarized differential decay rate can be expressed in terms of the 
structure functions:
\begin{eqnarray}
\frac{d^3\Gamma}{d\eta d\nu dq^2}=&&\frac{G_F^2|V_{qb}|^2}{32\pi^3E}
[W_1 2(q^2-M_\ell^2)+W_2(4\eta\nu-4\eta^2-q^2+M_\ell^2) \nonumber \\
&&+W_3 2(\nu q^2-2\eta q^2+M_\ell^2\nu)/M
+W_4 M_\ell^2(q^2-M_\ell^2)/M^2 \nonumber \\ 
&&+W_5 4M_\ell^2(\nu-\eta)/M]\, ,
\label{eq:any1}
\end{eqnarray}
where we define the Lorentz invariant $\eta\equiv k_\ell\cdot P/M$.
Equation (\ref{eq:any1}) shows that the $B$ meson structure functions  
could be separated and extracted
from the measurement of the differential decay rate at different values of
$\eta$ for fixed $\nu$ and $q^2$.

It would be useful to express the differential decay rate in terms of the 
angle $\theta$ between the charged lepton
and the $B$ meson in the virtual $W$ rest frame.  The angle
$\theta$ is related to the Lorentz invariants $\eta, \nu,$ and $q^2$ by
\begin{equation}
2\eta=(1+\frac{M_\ell^2}{q^2})\nu-(1-\frac{M_\ell^2}{q^2})\sqrt{\nu^2-q^2}
\,{\rm cos}\theta \, .
\label{eq:angle}
\end{equation}

Using Eq.~(\ref{eq:angle}), the differential decay rate in terms of 
${\rm cos}\theta$, $\nu$, and $q^2$ can be obtained from Eq.~(\ref{eq:any1}): 
\begin{eqnarray}
\frac{d^3\Gamma}{d{\rm cos}\theta d\nu dq^2}=&&\frac{G_F^2|V_{qb}|^2}
{64\pi^3E}(1-\frac{M_\ell^2}{q^2})\sqrt{\nu^2-q^2}\,[W_1 2(q^2-M_\ell^2)
+W_2 (\nu^2-q^2+M_\ell^2-\kappa^2) \nonumber \\
&&+W_3 2(q^2-M_\ell^2)\sqrt{\nu^2-q^2}\,{\rm cos}\theta/M+
W_4 M_\ell^2(q^2-M_\ell^2)/M^2 \nonumber \\
&&+W_5 2M_\ell^2(\nu-\kappa)/M]\, ,
\label{eq:any2}
\end{eqnarray}
where
\begin{equation}
\kappa=\frac{M_\ell^2}{q^2}\nu-(1-\frac{M_\ell^2}{q^2})\sqrt{\nu^2-q^2}
\,{\rm cos}\theta  .
\end{equation}
From Eq.~(\ref{eq:any2}) we see that the structure functions could also be
separated and extracted experimentally by looking at the angular $\theta$ 
distribution of the charged lepton at each value of $\nu$ and $q^2$. 

Integrating Eq.~(\ref{eq:any2}) over ${\rm cos}\theta$ gives
\begin{eqnarray}
\frac{d^2\Gamma}{d\nu dq^2}=&&\frac{G_F^2|V_{qb}|^2}{32\pi^3E}(1-\frac{
M_\ell^2}{q^2})\sqrt{\nu^2-q^2}\{ W_1 2(q^2-M_\ell^2)  \nonumber  \\
&&+W_2\frac{1}{3}[2\nu^2(1+\frac{M_\ell^2}{q^2}-\frac{2M_\ell^4}{q^4})
-2q^2+\frac{M_\ell^4}{q^2}-M_\ell^2]   \nonumber   \\
&&+W_4\frac{M_\ell^2}{M^2}(q^2-M_\ell^2)+W_5\frac{2M_\ell^2\nu}{M}(1-
\frac{M_\ell^2}{q^2})\}
\label{eq:any3}
\end{eqnarray}
This decay distribution would also be useful to measure the structure 
functions except for $W_3$, whose contribution is absent.

Equations (\ref{eq:any1}), (\ref{eq:any2}), and (\ref{eq:any3}) show that the 
contributions of the structure functions $W_4$ and $W_5$ are suppressed as
they are multiplied by the square of the charged lepton mass.  Therefore, the 
inclusive semileptonic $B$ decay with a tau lepton in the final state would 
be appropriate for measuring $W_4$ and $W_5$. 

In the inclusive semileptonic B decay where an electron (or muon) is the
final state lepton, the mass effect of it may be negligible.  Neglecting the
charged lepton mass, then
the differential decay rates in Eqs.~(\ref{eq:any1}), 
(\ref{eq:any2}), and (\ref{eq:any3}) reduce, respectively, to    
\begin{equation}
\frac{d^3\Gamma}{d\eta d\nu dq^2}=\frac{G_F^2|V_{qb}|^2}{32\pi^3E}
[W_1 2q^2+W_2(4\eta\nu-4\eta^2-q^2)+W_3\frac{2q^2}{M}(\nu-2\eta)]\, ,
\label{eq:zero1}
\end{equation}
\begin{equation}
\frac{d^3\Gamma}{d{\rm cos}\theta d\nu dq^2}=\frac{G_F^2|V_{qb}|^2}
{64\pi^3E}\sqrt{\nu^2-q^2}[W_1 2q^2+W_2 (\nu^2-q^2){\rm sin}^2\theta
+W_3\frac{2q^2}{M}\sqrt{\nu^2-q^2}\,{\rm cos}\theta]\, ,
\label{eq:zero2}
\end{equation}
\begin{equation}
\frac{d^2\Gamma}{d\nu dq^2}=\frac{G_F^2|V_{qb}|^2}{48\pi^3E}
\sqrt{\nu^2-q^2}[W_1 3q^2+W_2(\nu^2-q^2)]\, .
\label{eq:zero3}
\end{equation}
Note that only the structure functions $W_1$ and $W_2$ contribute to the double
differential decay rate $d^2\Gamma/(d\nu dq^2)$ with lepton masses being 
neglected.
 
The formulas for the differential decay rates derived in this section
are quite general and can be used to measure the $B$ meson structure functions
in any particular frame. However, we should point out that,
in addition to the lowest order contribution
in the weak interaction, there are higher order radiative processes which
contribute to the decay rate. In order to extract the structure functions 
from experimental data, the radiative corrections to the decay rate must be
implemented. The electroweak radiative corrections were discussed in
Ref.~\cite{ew}.
The QCD radiative corrections were calculated at the parton level of
quarks and gluons in Refs.~\cite{ali,ccm,jez,gremm}.  

\section{Behavior of the Structure Functions}
The behavior of the structure functions manifests the structure of the $B$
meson and the dynamics of strong interactions.
In this section we explore theoretically how the $B$ meson structure functions
look like.  We will also discuss the previous results in Ref.~\cite{jp}.  

In order to see the way in which the structure of the $B$ meson and strong
interactions manifest themselves in the structure functions, it will be useful
to start with an investigation of the behavior of the structure functions in 
the free
quark limit when no strong interaction takes place.  The unpolarized 
tree-level decay rate for the free point-like quark decay 
$b\to q\ell\bar\nu$ is given in arbitrary frame by
\begin{equation}
d\Gamma_{\rm free}=\frac{G_F^2|V_{qb}|^2}{(2\pi)^5p_b^0}L_{\mu\nu}
w^{\mu\nu}\frac{d^3k_\ell}{2E_\ell}\frac{d^3k_\nu}{2E_\nu}\, ,
\label{eq:free}
\end{equation}
where $p_b$ stands for the four-momentum of the $b$ quark of mass $m_b$. 
$L_{\mu\nu}$ is the leptonic tensor given in Eq.~(\ref{eq:klepton}). 
$w^{\mu\nu}$ is the quark tensor
\begin{equation}
w^{\mu\nu}=4\delta[(p_b-q)^2-m_q^2]\{-g^{\mu\nu}(m_b^2-q\cdot p_b)+
2p_b^\mu p_b^\nu+i\varepsilon^{\mu\nu\rho\sigma} p_{b\rho}q_\sigma-
(p_b^\mu q^\nu+q^\mu p_b^\nu)\}\, ,
\label{eq:qtensor}
\end{equation}
where $m_q$ is the mass of the decay produced $q$-quark.

In the free quark limit, the $B$ meson and the $b$ quark in it have the same 
velocity: $p_b/m_b=P/M$.  By use of the delta function relation
\begin{equation}
\delta[(p_b-q)^2-m_q^2]=\frac{1}{M(m_b-M\xi_-)}\delta(\xi_+-\frac{m_b}{M})\, ,
\end{equation}
where
\begin{equation}
\xi_\pm=\frac{\nu\pm\sqrt{\nu^2-q^2+m_q^2}}{M}\, ,
\end{equation}
comparing Eqs.~(\ref{eq:kga}) and (\ref{eq:exp2}) with 
Eqs.~(\ref{eq:free}) and (\ref{eq:qtensor}) allows us to identify 
the structure functions in the free quark limit
\begin{eqnarray}
W_1^{\rm free} & = & 2\delta(\xi_+-\frac{m_b}{M})\, , \label{eq:fw1}\\
W_2^{\rm free} & = & \frac{8\xi_+}{\xi_+-\xi_-}\delta(\xi_+-\frac{m_b}{M})\,,\\
W_3^{\rm free} & = & -\frac{4}{\xi_+-\xi_-}\delta(\xi_+-\frac{m_b}{M})\, ,\\
W_4^{\rm free} & = & 0\, ,\\
W_5^{\rm free} & = & W_3^{\rm free}\, . \label{eq:fw5}
\end{eqnarray}
The ``free'' functions $W_1^{\rm free}(\nu, q^2)/2$, $(\xi_+-\xi_-)W_2^{\rm 
free}(\nu, q^2)/8$, $(\xi_+-\xi_-)W_{3,5}^{\rm free}(\nu, q^2)/4$ exhibit the
intriguing property that they are functions of only one variable $\xi_+$ and 
not of $\nu$ and $q^2$ independently.  It is interesting to see if this 
scaling property persists under the strong interaction modification, 
which we shall discuss below.   

We go on to explore the actual behavior of the structure functions as 
functions of their arguments $\nu$ and $q^2$, switching on strong 
interactions.  In the inclusive semileptonic $B$ decays,
the square of the momentum transfer lies in the range $M_\ell^2\leq q^2
\leq (M-M_{X_{\rm min}})^2$, where $M_{X_{\rm min}}$ is the minimal hadronic 
invariant mass in the final state, which is the $D$ meson (pion) mass
for the $b\to c$ ($b\to u$) transition.  
Because of heaviness of the $B$ meson, the decays occur mostly at large 
momentum transfer.   
From Eq.~(\ref{eq:comm2}) we see that at large momentum transfer the hadronic 
tensor is dominated by the space-time region near the light cone $y^2\to 0$, 
where $y$ is the space-time
interval between the points at which the currents $j_\mu (y)$ and 
$j_\nu (0)$ act.  

The light-cone dominance implies that the five structure 
functions, {\it a priori} independent,
are related to a single distribution function \cite{jp,jin,jp1}:
\begin{eqnarray}
W_1 & = & 2[f(\xi_+) + f(\xi_-)]\, ,\label{eq:w1}\\
W_2 & = & \frac{8}{\xi_+ -\xi_-}[\xi_+f(\xi_+)-\xi_-f(\xi_-)]\,,\label{eq:w2}\\
W_3 & = & -\frac{4}{\xi_+-\xi_-} [f(\xi_+) -f(\xi_-)]\, ,\label{eq:w3}\\
W_4 & = & 0\, ,\label{eq:w4}\\
W_5 & = & W_3\, ,\label{eq:w5}
\end{eqnarray}
where the distribution function is defined by \cite{jp,jin,jp1}
\begin{equation}
f(\xi) = \frac{1}{4\pi M^2}\int d(y\cdot P)e^{i\xi y\cdot P}
\langle B|\bar{b}(0)P\!\!\!/(1-\gamma_5)b(y)
|B\rangle |_{y^2=0}\, .
\label{eq:distr3}
\end{equation}
The distribution function reflects the long-distance strong interaction 
responsible for the $b$-quark confinement in the $B$ meson.  
The structure functions show a simplified structure at large momentum 
transfer.  We see that on the light-cone dominance $W_4$ remains zero and 
also $W_5=W_3$ just as in the free quark limit.

It is instructive to pin down the form of the $b$ quark distribution 
function in the free quark limit for comparison with its form in the real
world.
In this limit, the free Dirac field $b(y)=e^{-iy\cdot p_b}b(0)$
with the velocity equality $p_b/m_b=P/M$, so from Eq.~(\ref{eq:distr3})
it follows that the distribution function becomes
\begin{equation}
f(\xi)=\delta(\xi-\frac{m_b}{M})\, .
\label{eq:disfree}
\end{equation}  
Substituting Eq.~(\ref{eq:disfree}) into Eqs.~(\ref{eq:w1})--(\ref{eq:w5}), we
recover the structure functions in the free quark limit in 
Eqs.~(\ref{eq:fw1})--(\ref{eq:fw5}).
Deviations from the delta function behavior would signal the strong
interaction effects.

Several important properties of the distribution function can be derived from 
field theory \cite{jp,jin1}.
It obeys positivity with a support $0\leq\xi\leq 1$.
The distribution function is normalized to unity:
\begin{equation}
\int_0^1 d\xi f(\xi)= 1 .
\label{eq:norm}
\end{equation}
The normalization of $f(\xi)$ is exact and does not get 
renormalized as a consequence of $b$ quantum number conservation.
Since the $b$ quark inside the $B$ meson behaves as almost free because of its
large mass, relative to which its
binding to the light constituents is weak, we expect
that the distribution function is very close to its asymptotic form --- the
delta function in Eq.~(\ref{eq:disfree}).  
This is supported by the analysis using the operator product expansion and 
the heavy quark effective theory method \cite{chay,bigi,manohar,blok,mannel},
which indicates that the distribution function is sharply peaked around $\xi=
m_b/M$ with a width of order $\Lambda_{QCD}/M$ \cite{jp,jin,jin1}.

It is convenient to redefine the structure functions as
\begin{eqnarray}
F_1(\nu, q^2) & = & \frac{1}{2}W_1(\nu, q^2)\,, \\
F_2(\nu, q^2) & = & \frac{\xi_+-\xi_-}{8}W_2(\nu, q^2)\,, \\
F_3(\nu, q^2) & = & \frac{\xi_+-\xi_-}{4}W_3(\nu, q^2)\,, \\
F_5(\nu, q^2) & = & \frac{\xi_+-\xi_-}{4}W_5(\nu, q^2)\,.
\end{eqnarray}
With these definitions, it is straightforward to get from 
Eqs.~(\ref{eq:w1})--(\ref{eq:w5}):
\begin{equation}
\frac{F_1-F_3}{2}=\frac{F_2+\xi_-F_1}{\xi_++\xi_-}=\frac{F_2+\xi_-F_3}
{\xi_+-\xi_-}=f(\xi_+)\, .
\label{eq:scaling1}
\end{equation}
Note that on the light-cone dominance $F_5=F_3$.  Equation (\ref{eq:scaling1})
shows scaling of the linear combinations of the structure functions, i.e., for 
large $q^2$ they depend only on the dimensionless scaling variable $\xi_+$.
The normalization condition (\ref{eq:norm}) of the distribution function 
implies the sum rules
\begin{equation}
\int_{\frac{m_q+M_\ell}{M}}^1 d\xi_+\, \frac{F_1-F_3}{2}=
\int_{\frac{m_q+M_\ell}{M}}^1 d\xi_+\, \frac{F_2+\xi_-F_1}{\xi_++\xi_-}=
\int_{\frac{m_q+M_\ell}{M}}^1 d\xi_+\, \frac{F_2+\xi_-F_3}{\xi_+-\xi_-}=1\,.
\label{eq:sumrule1}
\end{equation}

We may go further.  Assuming the $f(\xi_-)$ terms in 
Eqs.~(\ref{eq:w1})--(\ref{eq:w5}) to be negligible implies the following
relations of the structure functions \cite{jp} 
\begin{equation}
\xi_+F_1=F_2=-\xi_+F_3=-\xi_+F_5=\xi_+f(\xi_+)\, ,
\label{eq:scaling2}
\end{equation}
where the structure functions also scale, displaying the same property of the 
structure
functions as in the free quark limit shown above.  Furthermore, the following 
sum rules can be obtained \cite{jp} from the normalization 
condition (\ref{eq:norm}): 
\begin{equation}
\int_{\frac{m_q+M_\ell}{M}}^1 d\xi_+\, F_1=
\int_{\frac{m_q+M_\ell}{M}}^1 d\xi_+\, F_2/\xi_+=
-\int_{\frac{m_q+M_\ell}{M}}^1 d\xi_+\, F_3=
-\int_{\frac{m_q+M_\ell}{M}}^1 d\xi_+\, F_5=1\,.
\label{eq:sumrule2}
\end{equation}

The relations (\ref{eq:scaling1}) and (\ref{eq:scaling2}) are a consequence
of the spin $1/2$ nature of the $b$ quark; spin 0 partons would lead
to $F_1=0$ or $W_1=0$. 

The $f(\xi_-)$ term is a consequence of field theory, corresponding to the 
creation of quark-antiquark pairs in the final state.  
Here we pursue the arguments of \cite{jp} and examine more closely the 
assumption that the $f(\xi_-)$ terms in Eqs.~(\ref{eq:w1})--(\ref{eq:w5})
are negligible. The kinematic ranges of $\xi_+$ and $\xi_-$ are
\begin{eqnarray}
\frac{m_q+M_\ell}{M}\leq&\xi_+&\leq 1\, , \label{eq:xplus} \\
\frac{M_\ell-m_q}{M}(\frac{M_\ell+m_q}{M})^{\Theta(M_\ell-m_q)}\leq&\xi_-&
\leq 1-\frac{2m_q}{M}\, , \label{eq:xminus}
\end{eqnarray}
respectively,
where $\Theta(x)$ is a step function.
For $b\to c$ decays $\xi_-\lesssim 0.5$, the $f(\xi_-)$ terms
entering Eqs.~(\ref{eq:w1})--(\ref{eq:w5}) can be safely neglected since
the distribution function is sharply peaked near one.
However, for $b\to u$ decays $\xi_-$ can be as large as one, so that the sizes
of $f(\xi_-)$ and $f(\xi_+)$ may be comparable. 
Thus the approximate scaling and the sum rules in Eqs.~(\ref{eq:scaling2})
and (\ref{eq:sumrule2}) would hold for the inclusive charmed semileptonic
$B$ decay, but would be considerably violated by the non-negligible 
$f(\xi_-)$ term for the inclusive charmless semileptonic $B$ decay.  
By contrast, the approximate scaling and the sum rules in 
Eqs.~(\ref{eq:scaling1}) and (\ref{eq:sumrule1}) are valid regardless of
the size of $f(\xi_-)$.

Also worthy of noting is that there are other sources of scaling violation: 
logarithmic corrections due to gluon radiation and corrections of inverse 
powers of $q^2$ due to departures from the light cone.   

\section{Extraction of the distribution function}
The $b$ quark distribution function $f(\xi)$ defined in Eq.~(\ref{eq:distr3})
contains important information on the nature of confinement. As a necessary input,
its knowledge is also crucial for improving the theoretical accuracies on the 
determinations of $|V_{ub}|$ and $|V_{cb}|$ from the charged lepton energy
spectra \cite{jin,jp1} and the hadronic invariant mass spectrum \cite{jin2}. 
An experimental extraction of the distribution function will
therefore be of special importance and interest.  
In this section we shall discuss how
one could extract the distribution function from experiment.
 
Equations (\ref{eq:w1})--(\ref{eq:w5}) imply that $f(\xi)$ can be extracted
through the measurements of the structure functions, which is discussed in 
section II.  Recently,
it has been put forward \cite{jin3} that a measurement of the 
differential decay rate as a function of $\xi_+$ can be also used to extract 
directly the $b$ quark distribution function. 
The expression for $d\Gamma/d\xi_+$ has been derived in \cite{jin3},
neglecting lepton masses.  Here we generalize this result to the case where
the charged lepton mass cannot be neglected. 
Substituting Eqs.~(\ref{eq:w1})--(\ref{eq:w5}) into Eq.~(\ref{eq:any3}),
changing the variables from $(\nu, q^2)$ to $(\xi_+, q^2)$, and integrating
over $q^2$ with the kinematic limits $M_\ell^2\leq q^2\leq (\xi_+M-m_q)^2$, 
we arrive at
\begin{equation}
\frac{d\Gamma}{d\xi_+}=\frac{G_F^2|V_{qb}|^2}{192\pi^3}\frac{M^6}{E} \Bigg
[\xi_+^5f(\xi_+)\Phi(x, z)+\frac{12}{M^5}\int_{M_\ell^2}^{(\xi_+M-m_q)^2}
dq^2\sqrt{\nu^2-q^2}f(\xi_-)(1-\frac{M_\ell^2}{q^2})\Omega(\xi_+, q^2)\Bigg ]
\label{eq:disxi}
\end{equation}
with
\begin{eqnarray}
\Phi(x, z)=&&\sqrt{\lambda}[1-7(x^2+z^2+x^4+z^4+x^4z^2+x^2z^4)+12x^2z^2
+x^6+z^6] \nonumber \\
&&+12(2x^4-z^4-x^4z^4){\rm ln}\frac{1+x^2-z^2+\sqrt{\lambda}}{2x} \nonumber \\
&&+12(1-x^4)z^4{\rm ln}\frac{(1-x^2)^2-(1+x^2)z^2+(1-x^2)\sqrt{\lambda}}
{2xz^2} \, , 
\end{eqnarray}
\begin{equation}
\Omega(\xi_+, q^2)=q^2-M_\ell^2+\frac{2M_\ell^2\nu}{\xi_+M}(1-\frac{M_\ell^2}
{q^2})+\frac{1}{3}\frac{\xi_-}{\xi_+}[q^2+M_\ell^2-\frac{2M_\ell^4}{q^2}
-4\nu^2(1+\frac{M_\ell^2}{q^2}-\frac{2M_\ell^4}{q^4})] \, ,
\end{equation}
where 
\begin{equation}
\lambda=-4x^2+(1+x^2-z^2)^2\, ,\,\,\, x=\frac{m_q}{\xi_+M}\, , \,\,\, z=
\frac{M_\ell}{\xi_+M}\, ,
\end{equation}
$\nu$ and $\xi_-$ are related to $\xi_+$ and $q^2$ by
\begin{eqnarray}
\nu & = & \frac{\xi_+^2M^2+q^2-m_q^2}{2\xi_+M} \, ,\\
\xi_- & = & \frac{q^2-m_q^2}{\xi_+M^2} \, .
\end{eqnarray}
If the charged lepton mass is neglected, Eq.~(\ref{eq:disxi}) reduces 
to the expression given in \cite{jin3}.

The contribution of the $f(\xi_-)$ term to the specific spectra $d\Gamma/d\xi_+$
in Eq.~(\ref{eq:disxi}) is expected to be negligible
since (1) the distribution function is very close to its asymptotic form,
$\delta(\xi-m_b/M)$, in the free quark limit as discussed in the last 
section and in this limit $f(\xi_-)$ 
vanishes and (2) the dominant contribution to the $f(\xi_-)$ integral at
a given $\xi_+$ results from the large $\xi_-$ region, corresponding to
the neighbourhood of the upper integration limit for $q^2$, which is 
suppressed by $\nu^2-q^2$.  The numerical studies carried out in \cite{jin3}
have confirmed this expectation. 
The smallness of the $f(\xi_-)$ term has already been found in the parton
model \cite{parton}.  As a result,
$d\Gamma/d\xi_+$ is proportional to $f(\xi_+)$, so the measurement of 
the decay distribution
can be used to extract directly the nonperturbative distribution function.
In other words, the decay distribution with respect to the scaling variable
$\xi_+$ probes the QCD dynamics in a most sensitive manner. 

In practice,
as the measurements of the structure functions, radiative corrections must
be implemented in order to extract the distribution function from the measured
raw yields using the above expressions.

The decay distribution as a function of $\xi_+$ is a measurable quantity
with many other remarkable uses.  New hadronic uncertainty insensitive 
methods have been proposed \cite{jin3} for precise determinations of 
$|V_{ub}|$ and $|V_{cb}|$ through the measurements of the $b\to u$ and
$b\to c$ decay distributions as a function of $\xi_+$.  
Large theoretical uncertainties associated with nonperturbative strong
interactions can be avoided with these methods.
Moreover, in these measurements
rare $b\to u$ events can be separated efficiently from $b\to c$ events.   
These methods would enable us to determine especially $|V_{ub}|$ in a model 
independent way.   
The $b\to u$ spectrum $d\Gamma(B\to X_u\ell\nu)/d\xi_+$ could be also used to 
measure reliably the
inclusive charmless semileptonic branching fraction of the $B$ meson.
    
\section{Conclusions}
Inclusive semilpetonic $B$ decay is a very powerful means of probing hadron
structure and strong interactions. There are clear similarities between
inclusive semileptonic $B$ decay and deep inelastic lepton-nucleon scattering.
Inclusive semileptonic $B$ decay can be used to pursue the QCD as deep 
inelastic scattering.
We have presented general formulas for the differential decay rates for 
the inclusive semileptonic decays of the $B$ meson in terms of the structure
functions with the inclusion of charged lepton masses.
These formulas are suitable for measuring the $B$ meson structure functions 
in any particular frame.     

Measurements of the $B$ meson structure functions would provide  
evidence for the quark structure of the $B$ meson.
The scaling behavior of the structure functions at large momentum transfer
as in Eqs.~(\ref{eq:scaling1})
and (\ref{eq:scaling2}) would reveal the existence of a point-like $b$ quark 
in the $B$ meson.  The experimental verification of nonvanishing $W_1$ 
would indicate that the $b$ quark has spin $1/2$.
The sum rules as in Eqs.~(\ref{eq:sumrule1}) and (\ref{eq:sumrule2})
would provide a basic element in identifying the $b$ quantum number
of the $B$ meson.

The light-cone dominance simplifies the nonperturbative description of
inclusive semileptonic $B$ decays and permits 
the structure functions to be related to a single 
distribution function, leading to the testable theoretical predictions. 
The distribution function $f(\xi)$ is defined as the Fourier transform of the
matrix element of the bilocal and on-light-cone operator in $b$-quark
fields taken between the $B$ meson states, which gives the probability of 
finding a $b$ quark with momentum $\xi P$ inside the $B$ meson.  
This function depends only on the initial $B$ meson state and is therefore 
universal.  The nonperturbative distribution function
can be extracted by the measurements of either the structure functions or
the decay distribution as a function of $\xi_+$.  

In addition to testing the theoretical predictions discussed here,
advancing our understanding of hadron structure and strong
interactions, and providing new observations on the nature of confinement,
measurements of the $B$ meson structure functions and the universal $b$ quark 
distribution function would be beneficial to precise determinations of
the fundamental standard model parameters $|V_{ub}|$ and $|V_{cb}|$.     

The study of hadron structure and strong interactions by exploiting  
inclusive semileptonic $B$ decays will considerably extend the $B$ physics 
reach.  Measurements of the $B$ meson structure functions and the $b$ quark
distribution function should prove rewarding.
 
\acknowledgments
This work was supported by the Bundesministerium f\"ur Bildung, Wissenschaft,
Forschung und Technologie, Bonn, FRG under grant 057DO93P(7), and by the
Australian Research Council. 

{\tighten

} 

\end{document}